\documentclass[11pt]{article}
\usepackage{graphicx}
\usepackage{latexsym,graphicx}
\usepackage{float}
\usepackage{amsmath}
\usepackage{amsfonts}
\usepackage{amssymb}
\usepackage{epstopdf}
\usepackage{color}
\DeclareGraphicsExtensions{.eps}
\usepackage[left=2.5cm,top=2.5cm,right=2.5cm,bottom=2.5cm]{geometry}

\catcode `\@=11
\catcode `\@=12
\begin{document}
		\begin{center}
		\large{\bf{{The Reconstruction of Constant Jerk Parameter with f(R,T) Gravity}}} \\
		\vspace{5mm}
		\normalsize{Anirudh Pradhan$^1$, Gopikant Goswami$^2$, Aroonkumar  Beesham$^{4,5,6}$}\\
		\vspace{5mm}
		\normalsize{$^{1}$Centre for Cosmology, Astrophysics and Space Science (CCASS), GLA University, Mathura-281 406, Uttar Pradesh, India}\\
		\vspace{5mm}
		\normalsize{$^{2}$Department of Mathematics, Netaji Subhas University of Technology, Delhi, India}\\
		\vspace{5mm}
		\normalsize{$^{4}$Department of Mathematical Sciences, University of Zululand Private Bag X1001 Kwa-Dlangezwa 3886 South Africa}\\
		\vspace{5mm}
		\normalsize{$^{5}$National Institute for Theoretical and Computational Sciences (NITheCS), South Africa}\\
		\vspace{5mm}
		\normalsize{$^{6}$Faculty of Natural Sciences, Mangosuthu University of Technology, P O Box 12363, Jacobs 4052, South Africa}\\
		\vspace{2mm}
		$^1$E-mail: pradhan.anirudh@gmail.com \\
		\vspace{2mm}
		$^2$E-mail: gk.goswami9@gmail.com \\
	
		\vspace{2mm}
		$^{4,5,6}$E-mail: abeesham@yahoo.com \\
	\end{center}
	
	\begin{abstract}
		In this work, we have developed an FLRW type model of a universe which displays transition from deceleration in the past to the acceleration at the present. For this, we have considered field equations of f(R,T) gravity and have taken $f(R,T) = R + 2 \lambda T$, $\lambda$ being an arbitrary constant. We have estimated the $\lambda$ parameter in such a way that the transition red shift is found similar in the deceleration parameter, pressure and the equation of state parameter $\omega$. The present value of Hubble parameter is estimated on the basis of the  three types of observational data set: latest compilation of  $46$ Hubble  data set,  SNe Ia $580$ data sets of distance modulus and $66$ Pantheon data set of apparent magnitude which comprised of 40 SN Ia  bined and 26 high redshift data's in the  range $0.014 \leq z \leq 2.26 $. These data are compared with theoretical results through the $ \chi^2 $ statistical test. Interestingly, the model satisfies all the three  weak, strong and dominant energy conditions. The model fits well with observational findings. We have discussed some of the physical aspects of the model, in particular the age of the universe.
	\\
		
	\end{abstract}
	
		
\section{Introduction}  Cosmology as of now has become a very exciting and developing domain of knowledge in which we study how the universe originates(Hot Big bang)and  evolve over the time(Inflation, Era of subatomic heavy and light particles, radiation  and present day matter dominated Era) \cite{1} $-$ \cite{3}. Latest surveys and observations \cite{4} $-$ \cite{26}  
predict the  presence of  `Exotic and dynamic dark energy(DE) and opposite to it stationary 
cold dark matter(CDM) both in abundance. While CDM deviates the path of light and produces gravitational lensing, DE creates negative pressure and separates baryon galaxies and clusters 
from each other and creates acceleration in them. These two provide a larger observed value of critical density and distance modulus of luminous objects.\\

Before  these developments, the universe was best modeled by a well known standard FLRW spacetime which is spatially homogeneous and isotropic (Refer \cite{3} for details). The model originates with an unavoidable singularity representing bigbang. If the content of the universe can be assumed as a perfect fluid then it represents a decelerating expanding universe. This means that galaxies as fluid particles are moving away from one another with a relatively dampening pace over time.\\

In order to attribute acceleration in the FLRW model, cosmological constant $\Lambda$ was resurrected and  introduced in the Einstein field equation. It opined that $\Lambda$ supports the  acceleration $ \ddot{a}/a$ and opposes barion pressure so it is anti gravity and produces negative pressure. New cosmology is given the name $\Lambda$ CDM  (\cite{27} $-$ \cite{29}). Under this, it is proposed that nearly 27\% of the  total content of the universe is CDM and $\eqsim 70\%$ is dark energy which is interpreted as $\Lambda$ energy. $\Lambda$ CDM is also called a concordance model,
which is accepted by all the scientific community. But this model lacks in explaining fine tuning and cosmic coincidence problems\cite{30}. To avoid this, quintessence and phantom models \cite{31} $-$\cite{35} were developed in which  scalar field $\phi$ represents dark energy and satisfies the following equation of motion $\ddot{\phi}+3H\dot{\phi}+V'=0.$ It is very interesting that this equation is equivalent to energy conservation equation $\dot{\rho}+3H (p+\rho)=0$ for perfect fluid filled universe, where $\rho = \frac{1}{2}\dot{\phi}^2 + V(\phi)$ and  $ p = \frac{1}{2}\dot{\phi}^2 - V(\phi).$ Later on so many parametrizations were proposed for scalar $\phi$ and some interesting  cosmological models surfaced \cite{36} $-$ \cite{42}.\\

Parallelly, a group of scientists thought of a need to modify Einstein Field equations in order to explain the present day acceleration in the universe. For this, in the Einstein-Hilbert action $ S= \int(\frac{1}{16\pi G}( R + L_m)\sqrt{-g} dx^4,$ they replaced $R$ by  arbitrary functions of R, Trace of Einstein  $G_{ij}$ and  Energy momentum tensor $T_{ij}$. Accordingly  so many modified theories of gravity $ f(R)$, $ f(R,G) $, $ f(R,T) $, $ f(R,T^{\phi}) $, $f(Q)$, $f(T, B)$ and many more were proposed (\cite{43}$-$ \cite{76}). The idea behind this is as follows. Einstein's basic philosophy lies in the fact that matter and energy are equivalent and their presence in the universe being permanent in nature creates curvature in the spacetime representing it. In the Einstein field equations, the Ricci scalar $R$ represent curvature. In FLRW spacetime scale factor and its derivatives are related functionally with $R$. So by replacing it with some arbitrary non-linear function of R and T, we must see the change in the mechanical equations of General Relativity and we might get the universe accelerating instead of deceleration.\\

In this work, we tried to obtain an universe model based on the cosmological principle that the physical  universe is spatially homogeneous and isotropic and  that it displays transition from deceleration in the past to the acceleration at the present. For this, we solve $f(R,T)$ gravity field equations in the background of FLRW space time which represent a spatially homogeneous and isotropic universe. In order to get the desired result we have considered the simplest form of $f(R,T)$ as  $f(R,T) = R + 2 \lambda T$, where $\lambda$ is an arbitrary constant. While working with a universe model, one has to take care of the latest observational findings \cite{20} $-$ \cite{23}. In this contest estimations of present values of cosmological parameters like Hubble $H_0$, deceleration parameter $q_0$, growth of density, transition behaviour of pressure and equation of state parameter become crucial. Seeing the complex structure of the universe which contains so many exotic energies like dark energy in abundance, dark matter, black holes etc, roll of higher order derivative para meters like jerk, snap and lurk parameters become also crucial. Our universe at large scale is very much discipline. This is the reason that a very simple $ \Lambda$CDM concordance model fits well on observational ground. In this model jerk parameter $j =1 $. In this paper too, we have been  successful to develop a  universe model in $f(R,T)$ gravity with constant jerk parameter. The interesting part in the paper is that  $\lambda$ parameter is chosen in such a way that the transition red shift is found similar in deceleration parameter, pressure and the equation of state parameter $\omega$ and that the  model satisfies all the three  weak, strong and Dominant energy conditions. The present value of Hubble parameter is estimated on the basis of the  three types of observational data set: latest compilation of  $46$ Hubble  data set,  SNe Ia $580$ data sets of distance modulus and   66 Pantheon data set of apparent magnitude which comprised of 40 SN Ia  binned and 26 high redshift data's in the  range $0.014 \leq z \leq 2.26 $. These data are compared with theoretical results through the $ \chi^2 $ statistical test. The model fits well with observational findings. We have discussed some of the physical aspects of the model in particular age of the universe.\\

The paper is presented in the following section wise form. In section $2$, the $f(R,T)$ field equations have been presented and are solved for spatially homogeneous and isotropic FLRW space time by taking the content in the universe as perfect fluid and accordingly we have taken energy momentum tensor as that of perfect fluid. We arrive at the two equations in which first one may be called equation of motion of a fluid particle as it contains second order derivative of scale factor in form of deceleration constant and the pressure term in the right hand side of the equation. The other second equation provides rate of expansion as it contains Hubble parameter and density term in the right hand side of the equation. In this section we have solved density, pressure and equation of state parameter $\omega$ in term of Hubble and decelerating parameter so that they can be obtained once we know about the Hubble and decelerating parameter. In section $3$, deceleration  and Hubble parameter were obtained by taking  jerk parameter constant. It is interesting that decelerating parameter shows transition from negative to positive over red shift which means that in the past universe was decelerating and at present it is accelerating. In section $4$,  density, pressure and the equation of state parameter $\omega$ were solved. The latter to show transition from  present to past. We have estimated $\lambda$ parameter in such a way that the transition red shift is found similar in deceleration parameter, pressure and the equation of state parameter $\omega$. Energy conditions are evaluated and discussed in section $5$. Interestingly, the model satisfies all the three  weak, strong and Dominant energy conditions. In section $6$, we have formulated distance modulus and apparent magnitude for our model and estimation of present value of Hubble parameter is done. As stated earlier, we have used the three types of observational data set: latest compilation of  $46$ Hubble  data set,  SNe Ia $580$ data sets of distance modulus and $66$ Pantheon data set of apparent magnitude which comprised of $40$ SN Ia  binned and $26$ high redshift data's in the  range $0.014 \leq z \leq 2.26 $. Figures and a table are properly presented to describe and analyse the results more clearly................ In the last Section $8$, conclusions are provided.
\section{f(R,T) gravity field equations for FLRW flat spacetime:}
As it is said in the introduction that in the $f(R,T)$ gravity \cite{56}, the Ricci scar $R$ is replaced by an arbitrary function f(R,T) of $R$ and $T$, so we write the two actions and  field equations as follows:\\

Einstein-Hilbert action:
\begin{equation}{\label{1}}
	S= \int(\frac{1}{16\pi G}( R+2\lambda) + L_m)\sqrt{-g} dx^4.
\end{equation}
Einstein GR field equations:
\begin{equation}\label{2}
	R_{ij}-\frac{1}{2} R g_{ij}+ \Lambda g_{ij} = \frac{8\pi G}{c^4}T_{ij}.
\end{equation}
f(R,T) gravity action:
\begin{equation}{\label{3}}
	S= \int\Big(\frac{1}{16\pi G} f(R,T)+L_m\Big)\sqrt{-g} dx^4,
\end{equation}
and
f(R,T) gravity field equations:
\begin{equation} {\label{4}}
	R_{ij}-\frac{1}{2} R g_{ij} = \frac{8 \pi G T_{ij}}{f^R(R,T)}+ \frac{1}{f^R (R,T)} \bigg(\frac{1}{2} g_{ij} (f(R,T)-R f^R (R,T)) - (g_{ij} \Box - \nabla_i \nabla_j) f^R (R,T) + f^T (R,T) (T_{ij} +p g_{ij}) \bigg),
\end{equation}
where the energy momentum tensor $T_{ij}$ is related to the matter Lagrangian $L_m$ via the following equation
\begin{equation}{\label{5}}
	T_{ij}= -\frac{2}{\sqrt{-g}}\frac{\delta(\sqrt{-g} L_m)}{\delta g^{ij}}.
\end{equation}
and  $ f^R $ and $ f^T $ are the derivative of $ f(R,T) $ with respect to $ R $ and $ T $ respectively. Three particular functional  forms for arbitrary function  $ f(R,T) $ have been proposed for cosmology \cite{56}. They are (a:) $R + 2f(T) $ (b:) $ f_{1}(R) + f_{2}(T)$ and (c:) $ f_{1}(R) + f_{2}(R)f_{3}(T).$\\

We solve  $ f(R,T) $ gravity field equations (\ref{4} ) for FLRW spatially flat spacetime given by

\begin{equation}{\label{6}}
	ds^{2} = dt^{2} - a^{2}(t) (dx^{2} + dy^{2} + dz^{2}),
\end{equation}
by taking  $ L_m = - p $ and  $ f(R,T) = R + 2\lambda T $  so that we get $T_{ij}$ as
\begin{equation}\label{7}
	T_{ij}= (\rho + p) u_i u_j - p g_{ij}.
\end{equation}
and
\begin{equation}{\label{8}}
	2 \dot{H} + 3 H^2 = -(8 \pi + 3 \lambda) p + \lambda \rho
\end{equation}
and
\begin{equation}{\label{9}}
	3 H^2 = (8 \pi + 3 \lambda ) \rho - \lambda p,
\end{equation}
where $H=\frac{\dot{a}}{a}$ is the Hubble parameter, $u^{i} u_{i} =1$ and $u^{i} u_{i;j} =0$.
We  assume $ \lambda = 8 \pi \mu $. Then the field equations (\ref{8}) and (\ref{9}) are simplified as:

\begin{equation}{\label{10}}
	(1 - 2 q) H^2 = 8 \pi ( \mu \rho - (1 + 3 \mu) p  )
\end{equation}
and
\begin{equation}{\label{11}}
	3 H^2 = 8 \pi  ( ( 1 + 3 \mu ) \rho - \mu p ),
\end{equation}
where $q=-\frac{\ddot{a}}{a H^2}$ is deceleration parameter and $ \dot{H} = - (q+1)H^2 $.
Eqs. (10) and (11) are the two fundamental equations which describe equation of motion of a fluid particle which is commonly a galaxy and they do describe rate of expansion of the universe(Hubble parameter). As is mentioned in the introduction, our universe is accelerating so the two equations must explain it. For this both deceleration constant and the pressure must be negative. The observations tell us that luminous content of the universe (baryon fluid) is at present dust so the baryon pressure must be zero. But the literature \cite{30} says that apart from baryon matter other energies do exist in the universe. It is estimated that nearly 28 \% of the total
content of the universe is dark matter which is responsible for the phenomenon of gravitating lensing occurring in the universe and nearly 68\% of energy exists in the form of dark energy which is responsible for the present day acceleration in the universe. These ideas and how to accommodate them in the theories, have been explained in the introduction. In $f(R,T)$ gravity theory, Ricci scalar $R$ is replaced by an arbitrary function of $R$ and trace of energy momentum tensor $T$. The idea is that we must expect acceleration due to curvature and trace dominance. The authors feel that the pressure term arising in the field equations is not due to the baryonic content of the
universe. It is as a result of the over all effect. We mean that terms containing $\mu$ in the field equations  (\ref{10}) and (\ref{11}) are the extra terms in the original FLRW field equations of general relativity and they will have impact in producing pressure and creates acceleration in the universe.
From Eqs. (\ref{10}) and (\ref{11}), we get, $\rho$, p and equation of state parameter $\omega = \frac{p}{\rho} $ as follows:
\begin{equation}\label{12}
8\pi	\rho =	\frac{H^2 (2 \mu  (q+4)+3)}{(2 \mu +1) (4 \mu +1)}
\end{equation}

\begin{equation}\label{13}
8\pi	p = \frac{H^2 ((6 \mu +2) q-1)}{(2 \mu +1) (4 \mu +1)}
\end{equation}
and 
\begin{equation}\label{14}
	\omega=\frac{p}{\rho}=\frac{(6 \mu +2) q-1}{2 \mu  (q+4)+3}
\end{equation}


\section{Model with Constant Jerk and Transition Deceleration Parameters and their Evaluation from OHD Data Set: }
There are two more parameters jerk and snap which are related with third and forth order of derivatives of scale factor. They play very important role in examining instability of a cosmological model. jerk is also one of the parameter in statefinder diagnostic. they are defined as: jerk $j=\frac{\dddot{a}}{a H^3}$ and snap $s= -\frac{\ddddot{a}}{a H^4}. $
The jerk parameter in the terms of deceleration parameter can be written as: 
\begin{equation}{\label{15}}
	j(z) = q(z) +2 {q(z)}^2 + (1+z) \frac{d q(z)}{d z}.
\end{equation}
As jerk parameter '$j$' varies very slowly and its present value lies around 1 ( j = $1$ as per $\Lambda$CDM model), so we solve Eq.(\ref{15}) for constant jerk i.e. we take j = constant.
We obtain following expression for deceleration parameter.
\begin{equation}\label{16}
q =	\frac{1}{4}\left(-1+\sqrt{-8 j-1}\text{Tan}\left[\frac{1}{2}\left(-2 \tan ^{-1}\left(\frac{4 \sqrt{-8 j-1} q_0+\sqrt{-8 j-1}}{8 j+1}\right)-\sqrt{-8 j-1} \log (z+1)\right).\right]\right),
\end{equation} 
where we have used $q = q_0~ \text{at~ present}~ i.e.~ z=0.$
The deceleration parameter $q$ is related to Hubble parameter through the following differential equation.
\begin{equation}\label{17}
	H_z(1+z)=(q+1)H	
\end{equation}
so, using  Eq. (\ref{16}) and integrating Eq. (\ref{17}), we may get the  expression for Hubble parameters, 
\begin{equation}\label{18}
	H = H_0 e^{\int_{0}^{z}\frac{(q+1)dz}{(1+z)}}.
\end{equation}
We note that the Eq. (\ref{18}) has three unknown parameters $H_0,~ j~ \text{and}~ q_0.$

Now, we will estimate present value of Hubble parameter $H_0$. This will be done statistically. For this we consider a data set of $77$ observed values of Hubble constant at different red shift using (a). Cosmic chronometric method, (b). BAO signal in galaxy distribution and (c). BAO signal in Ly$\alpha$ forest distribution alone or cross-correlated with QSOs \cite{77}$-$\cite{91}. This data set is commonly known as OHD data set. We compare the data set values from those obtained theoretically from Eq. (\ref{18}) by forming the following chi squire function of the present value of $H_0,~ j~ \text{and}~ q_0.$
\begin{equation}\label{18a}
	\chi^{2}( H_0,j,q_0) =\frac{1}{46} \sum\limits_{i=1}^{46}\frac{[Hth(z_{i},H_0,j,q_0) - H_{ob}(z_{i})]^{2}}{\sigma {(z_{i})}^{2}},
\end{equation}
Now the parameters $H_0,~ j~ \text{and}~ q_0.$ are estimated by getting the minimum value of chi squire for the values of $ H_0,~ j~ \text{and}~ q_0.$ taken in the range (65$-$75), (0.95$-$1.05) and(-0.60~$-$~-0.45) respectively. It is found that  $ H_0 = 68.95,~j=0.95~ \text{and}~ q_0 = -0.57$ , for minimum chi squire $\chi2$ =0.5378. It may be be called a good fit. The fit will be more clear from the error bar and livelihood plots in the figure $1$. Before that, we rewrite Eq. (\ref{16}) and integrate Eq. (\ref{18}) by taking $q_0=-0.57$ and $j=0.95$ which are estimated values as per OHD data set. We get the following expressions for deceleration and Hubble parameters as:
 
\begin{equation}\label{18b}
q = \frac{0.483144 \left( (z+1)^{2.93258}-5.18713\right)}{(z+1)^{2.93258}+2.5491}	
\end{equation}
and
\begin{equation}\label{18c}
	H = 0.54 H_0 (1+z)^{0.017}(3.55 + z(3+z(3+z)))^{0.489}
\end{equation}
\\
Figures $1(a)$ and $1(b)$ describe the growth of Hubble parameter $H$ and  expansion rate $H/(1+z) = \dot{a}/a_0$ over red shift $ `z'$ respectively. Hubble parameter is increasing function of red shift which means that in the past Hubble constant was more. It is gradually decreasing over time. Expansion is high at present which indicates that universe is accelerating. These figures also show that theoretical graph passes near by through the dots which are observed values at different red shifts. Vertical lines are error bars. Figure $5(c)$ is likelihood probability curve for Hubble parameter. Estimated value $H_0=68.95$ is at the peak.

\begin{figure}[H]
	(a)	\includegraphics[width=9cm,height=8cm,angle=0]{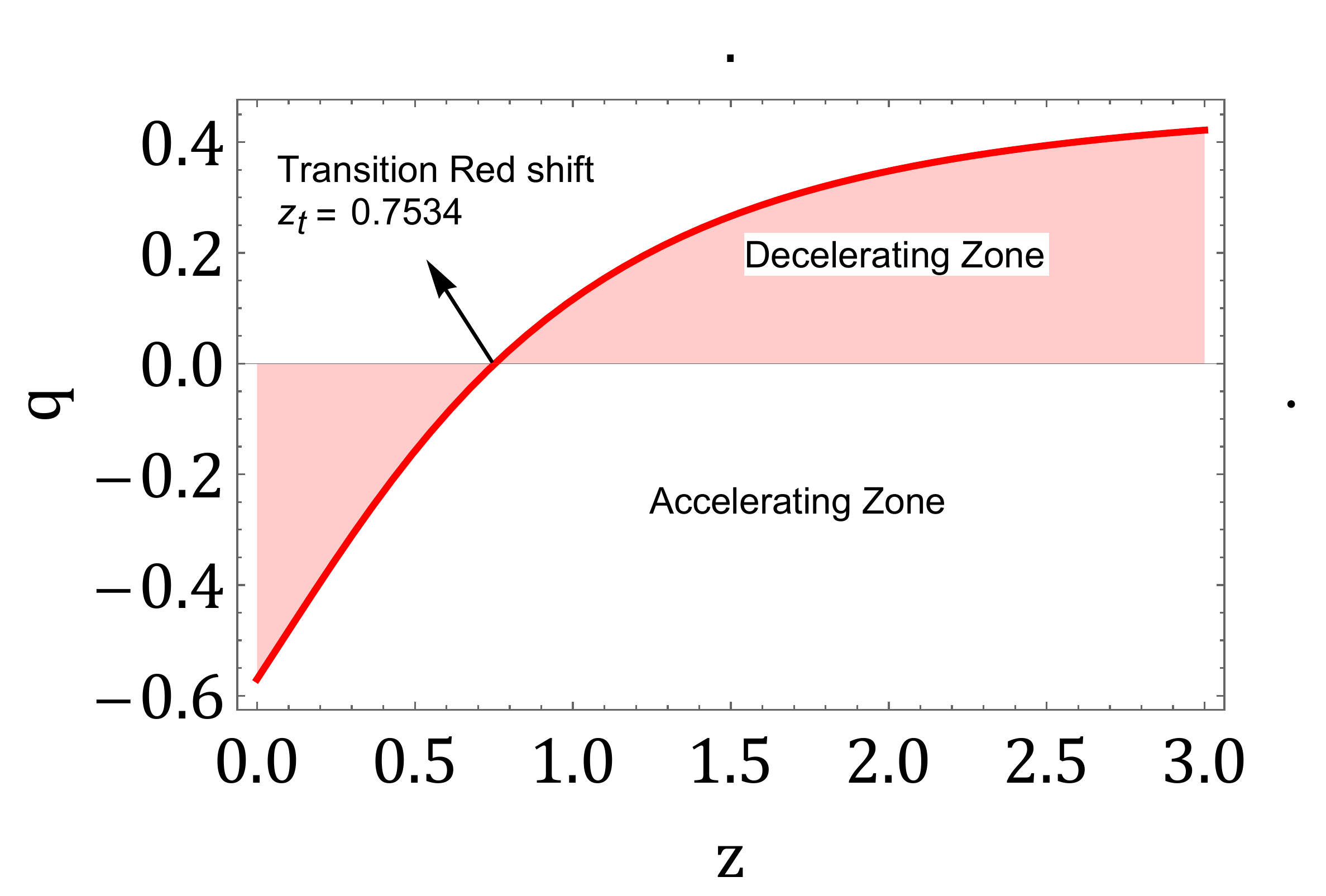}
	(b)	\includegraphics[width=9cm,height=8cm,angle=0]{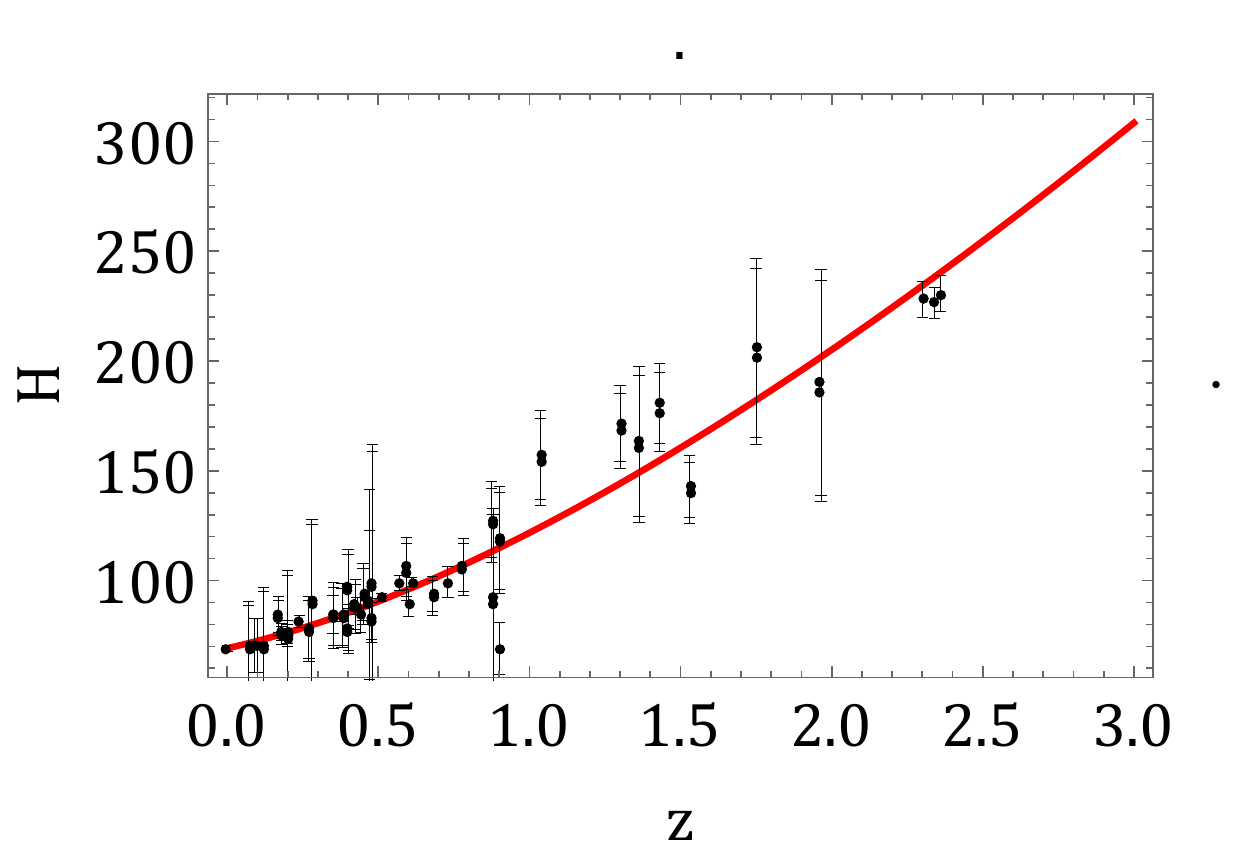}
	(c) \includegraphics[width=9cm,height=8cm,angle=0]{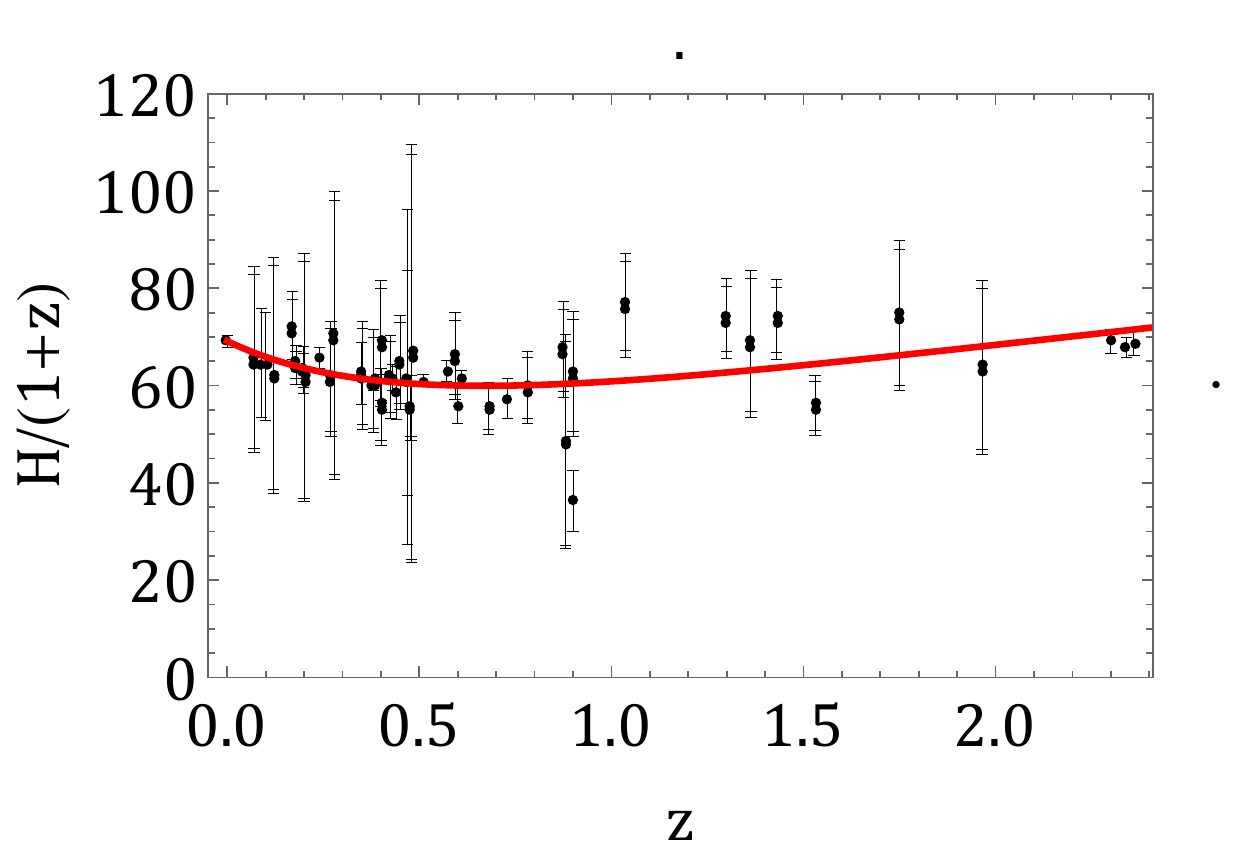}
	(d) \includegraphics[width=9cm,height=8cm,angle=0]{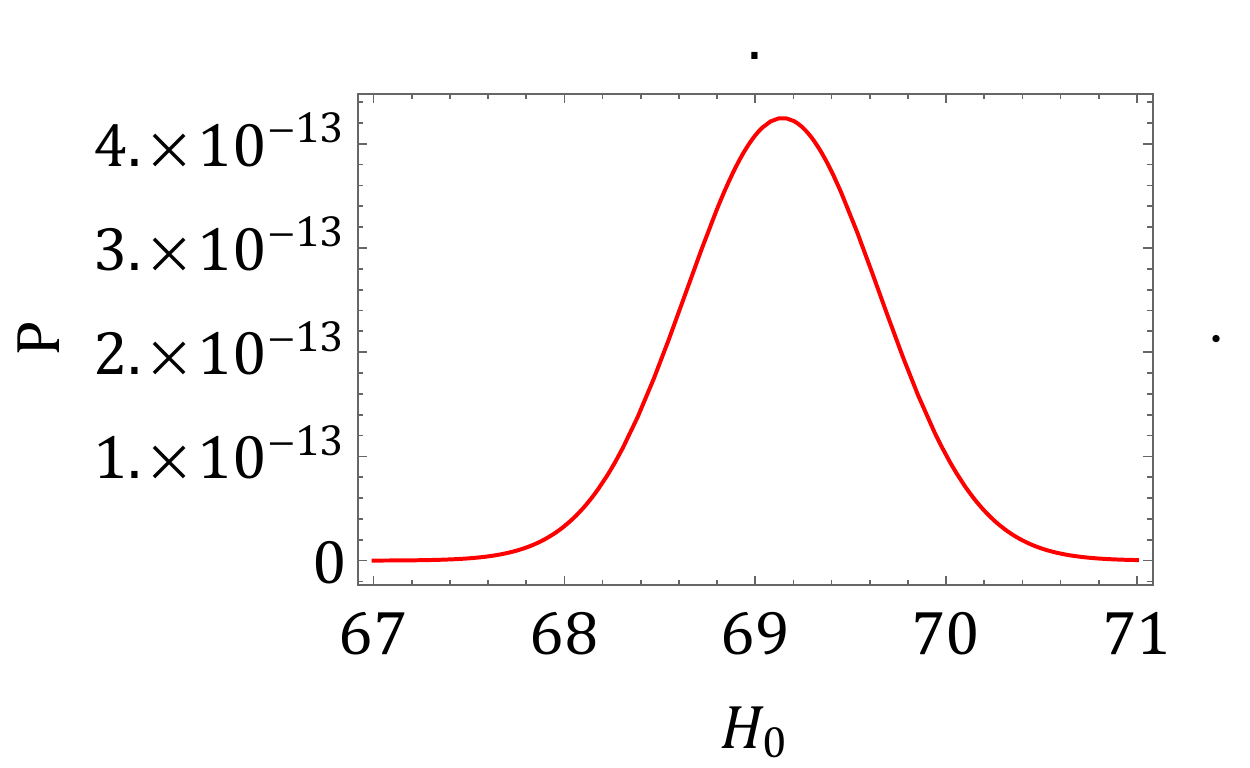}
	\caption{ Figure (a) shows the growth of deceleration parameter $`q'$  over red shift $ `z'$. It describes that in the past the universe was decelerating. At transition redshift $z_t$=0.7534, where $q\sim0$, it changed its behaviour and started accelerating.  Here $q_0 =-0.57~\text{and}~ j = 0.95.$  Figures (b) and (c) are the error bar plots  for  Hubble parameter $H$ and  expansion rate $H/(1+z) = \dot{a}/a_0$ over red shift $ `z'$ respectively. Figure (d) is likelihood probability curve for Hubble parameter. Estimated value $H_0=68.95$ is at the peak.}
\end{figure}

\section{Expressions for Density, Pressure and Equation of State parameter of the Universe:}
Having obtained expressions for deceleration, Hubble parameters and the value of $\mu$, we can obtain expressions for density $\rho$, pressure $p$ and equation of state parameter $\omega $from Eqs. (\ref{12})-(\ref{14}) which are given as below:

\begin{equation}\label{19}
	\rho = \rho_0\frac{0.2916 (z+1)^{0.034} (z (z (z+3)+3)+3.55)^{0.978} \left(2 \mu  \left(\frac{0.483144 \left((z+1)^{2.93258}-5.18713\right)}{(z+1)^{2.93258}+2.54909}+4\right)+3\right)}{6.86 \mu +3},
\end{equation}
\begin{equation}\label{20}
	p =\text{p0}\frac{0.2916  (z+1)^{0.034} (z (z (z+3)+3)+3.55)^{0.978} \left(\frac{0.483144 (6 \mu +2) \left((z+1)^{2.93258}-5.18713\right)}{(z+1)^{2.93258}+2.54909}-1\right)}{(6 \mu +2) \text{q0}-1}
\end{equation}
and
\begin{equation}\label{21}
	\omega =\frac{\mu  \left(0.323307 (z+1)^{2.93258}-1.67704\right)-0.00375986 (z+1)^{2.93258}-0.843309}{1.71537 \mu +1. \mu  (z+1)^{2.93258}+0.334587 (z+1)^{2.93258}+0.852892}
\end{equation}
\\{\LARGE }
From Eq. (\ref{21}), $\omega_0$ can be obtained from the parameter $\mu$ and $q_0$ as follows.
\begin{equation}\label{22}
	\omega_0 = -\frac{0.094}{\mu +0.44}-0.499,
\end{equation}
\\
We see here that as universe is accelerating at present, so the pressure should be negative and as such the equation of state  parameter  $\omega_0$($p=\omega \rho)$ should also be negative, As per $\Lambda$ CDM and quintessence dark energy  models   $\omega_0=-1$ and $\omega_0\leq -1$. Like deceleration parameter $`q'$,  Eq. (\ref{21}) also describe transition of equation of state parameter $\omega$ from negative values to positive values over red shift $`z'$. This will make pressure of the universe also transitional which means that in the past pressure was positive and at present it is negative which creates acceleration in the universe. It is very interesting to see that $\omega = -0.0008$ at $z= 0.7534$ for the value of parameter $\mu$ = 130. We note transition red shift for deceleration parameter $`q'$ is also $z_t$=0.7534. So the traditional red shift for both deceleration parameter and equation of state parameter matches for  $\mu$ = 130. This is very well seen in the following figure $2$ for $\omega$ and pressure. For drawing plot of pressure, we have replaced $p_0$ by $\omega_0 \rho_0$. The value of $\omega_0$ is obtained from Eq. (\ref{22}) as $\omega_0$=-0.499264 by taking $\mu$ = 130. The value of $\rho_0$ is taken from literature as $1.88\times 10^{-29} gm/cm^3$. The unit for pressure is taken as $dyne/cm^2$ in C.G.S unit.\\
The following figure $2$ describes that in the past both the  equation of state parameter $\omega$ and pressure were positive which means that universe was deceleration. At transition redshift $z_t$= 0.7534, where $\omega\sim 0 ~\&~ p \sim 0$, they changed their behaviour and become negative which means that universe started accelerating at this junction.

\begin{figure}[H]
	(a)	\includegraphics[width=9cm,height=8cm,angle=0]{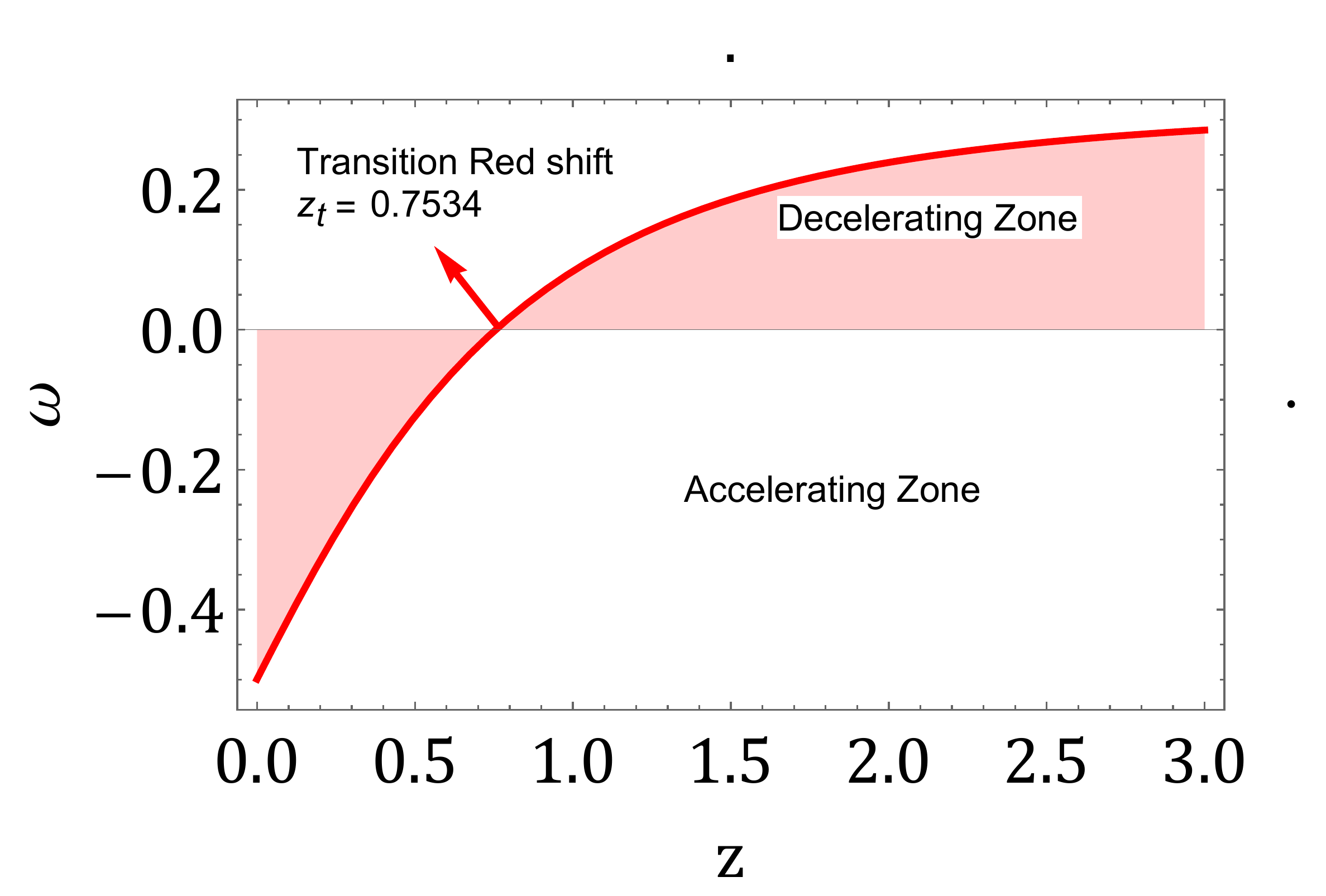}
	(b) \includegraphics[width=9cm,height=8cm,angle=0]{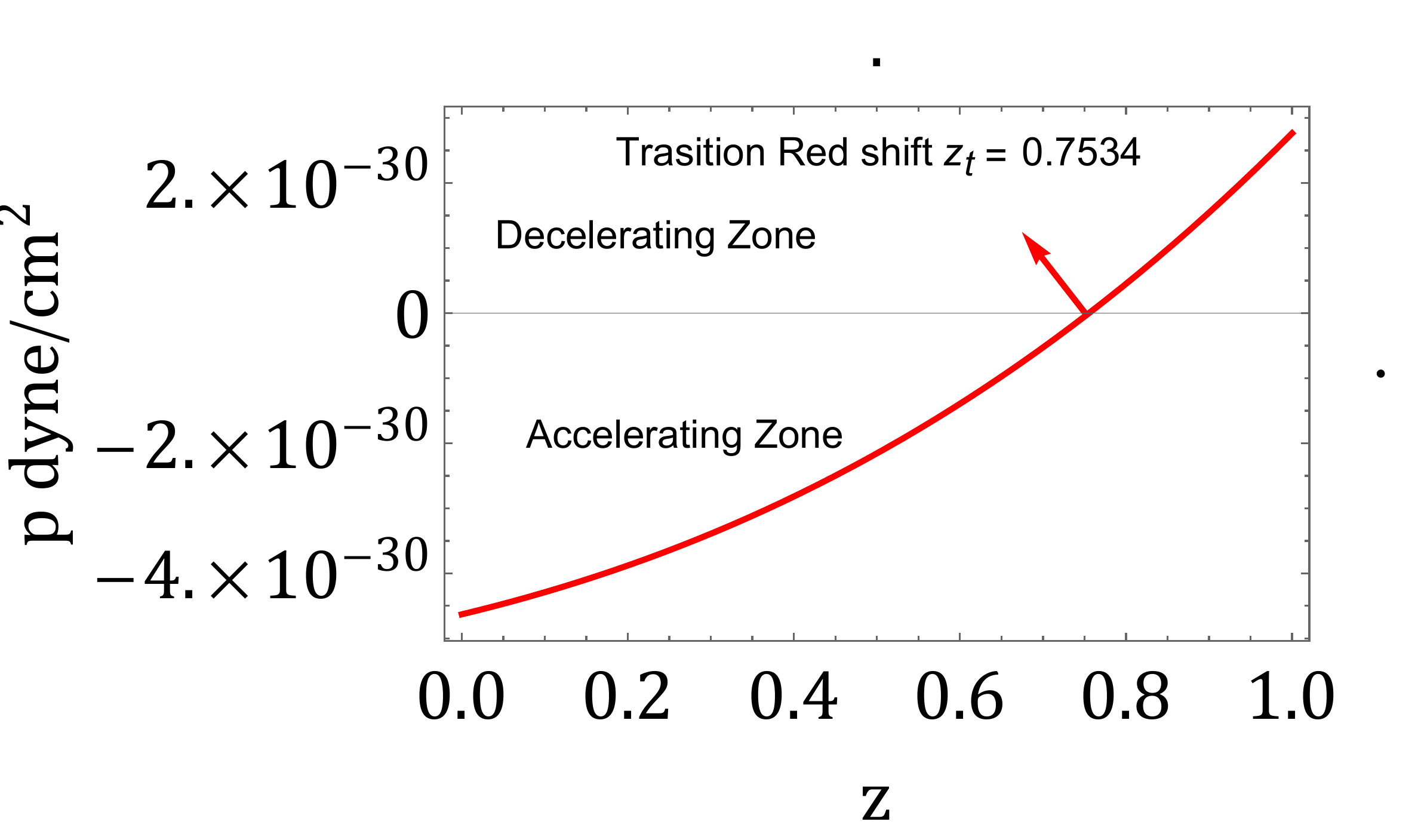}
	\caption{Figures (a) and (b) show the growth of equation of state parameter
		$\omega$ and pressure over red shift $ `z'$ respectively. It describes that in the past both the  equation of state parameter $\omega$ and pressure were positive which means that universe was deceleration. At transition redshift $z_t$= 0.7534, where $\omega\sim 0~ \& ~ p \sim 0$, they changed their behaviour and become negative which means that universe started accelerating at this junction.  Here $\mu = 130.$}
\end{figure}
We can also plot for density of the universe from Eq. (\ref{12}). The following figure $3$ describes that in the past density was more. It gradually decreases due to expansion of the universe. As we describe earlier, $\rho_0$ is taken  as $1.88\times 10^{-29} gm/cm^3$.
\begin{figure}[H]
	\centering
	\includegraphics[width=9cm,height=8cm,angle=0]{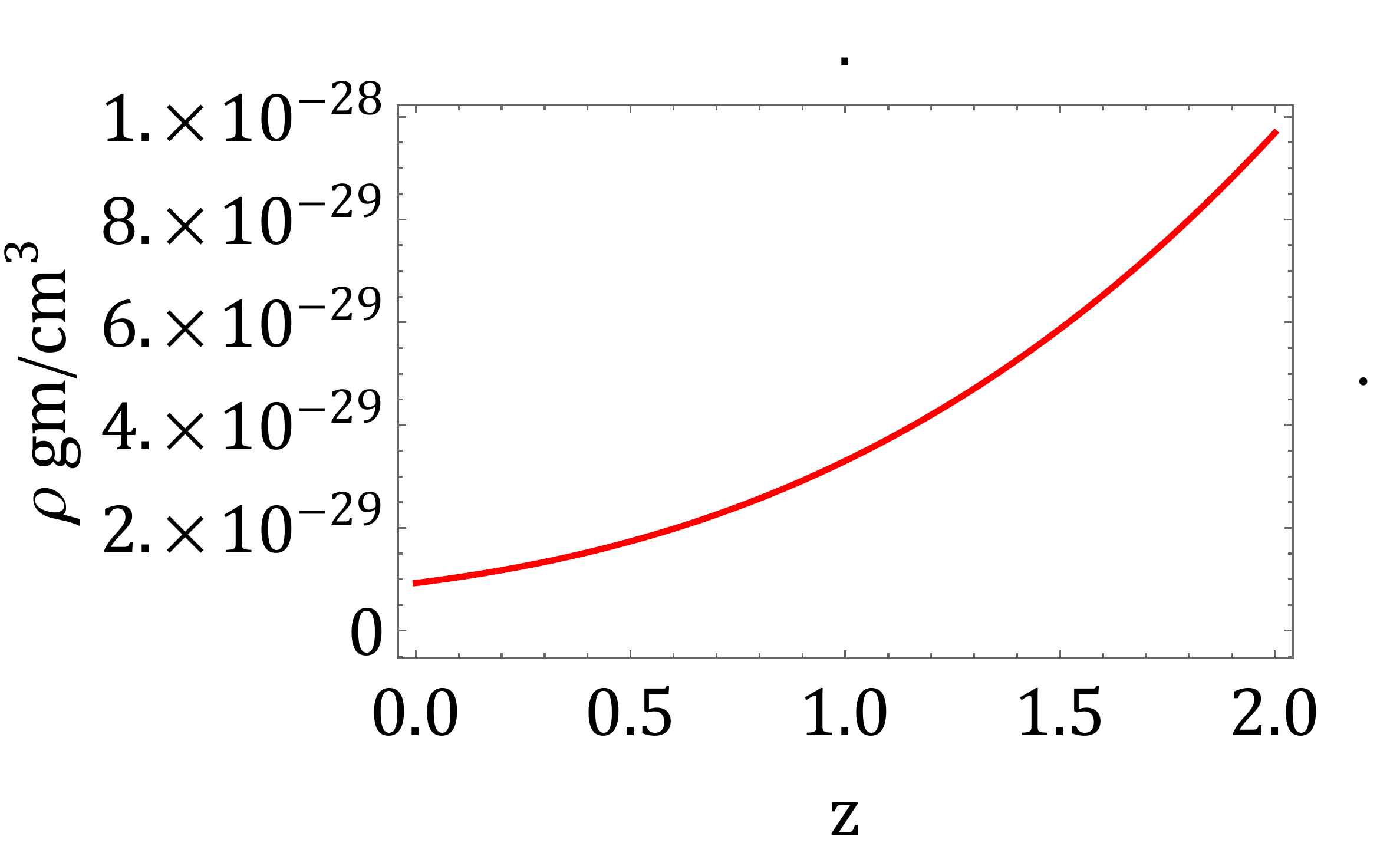}
	\caption{Figure describes the growth of density over red shift which is a increasing function. This means that in the past density was more. It gradually decreases due to expansion of the universe. Here $\mu = 130.$}
\end{figure}
\section{Energy Conditions}
There are the three types of energy conditions \cite{95} (a) Weak energy condition(WEC) $p+\rho \geq 0$ (b) Strong energy condition(SEC) $3p+\rho \geq 0 $ and (c) Dominant energy condition(DEC) $\rho -p \geq 0$. Apart from these, the common condition in all the three is that the density should be positive (i.e. $ \rho \geq 0 $).
The expressions for the above energy condition are as follows:
\begin{equation}\label{23}
	p(z)+\rho (z) =\text{$\rho $0} \frac{0.0145873  \left(34.5254 (z+1)^{2.96658}+1 (z+1)^{0.034}\right) (z (z (z+3)+3)+3.55)^{0.978}}{(z+1)^{2.93258}+2.54909}
\end{equation}

\begin{equation}\label{24}
	3~ p(z)+\rho (z) = \text{$\rho $0}\frac{1.2644  \left( -(z+1)^{0.034}+0.592556 (z+1)^{2.96658}\right) (z (z (z+3)+3)+3.55)^{0.978}}{(z+1)^{2.93258}+2.54909}
\end{equation}
and

\begin{equation}\label{25}
	\rho (z) - p(z) =\text{$\rho $0}\frac{1.29358  \left(0.199473 (z+1)^{2.96658}+1. (z+1)^{0.034}\right) (z (z (z+3)+3)+3.55)^{0.978}}{(z+1)^{2.93258}+2.54909}.
\end{equation}
It is interesting that  all the three conditions are being satisfied here despite negative pressure at present. It is more clear from the following figure $4$.
\begin{figure}[H]
	(a)	\includegraphics[width=9cm,height=8cm,angle=0]{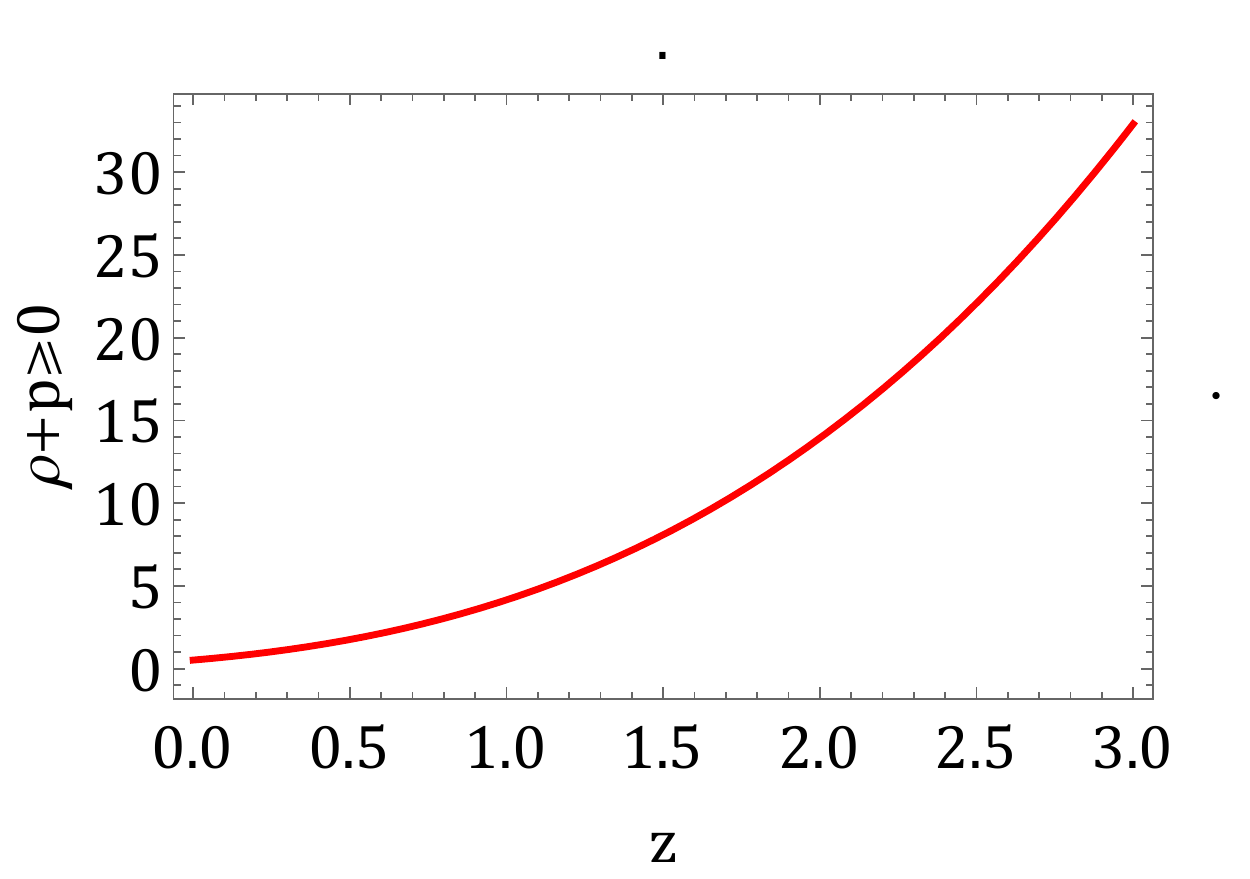}
	(b) \includegraphics[width=9cm,height=8cm,angle=0]{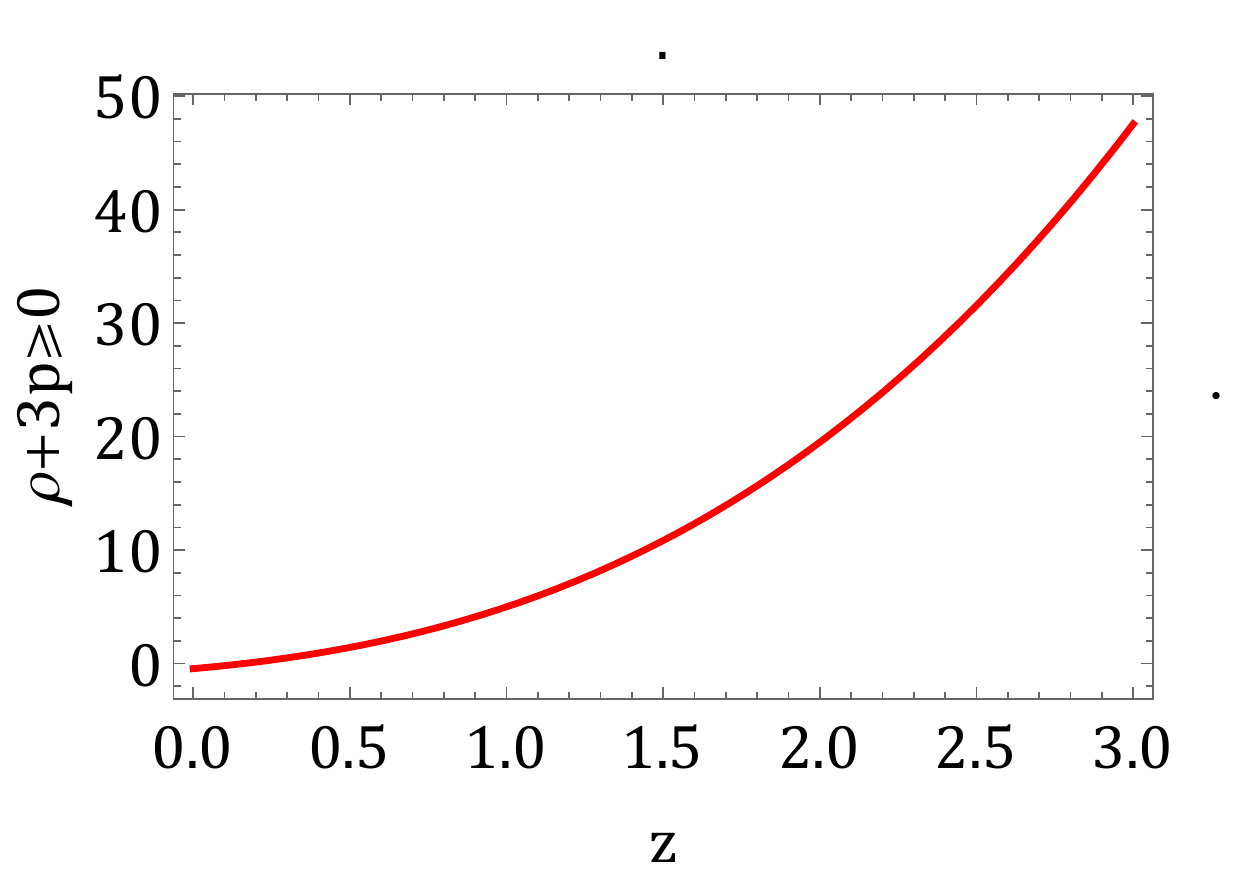}
	(c) \includegraphics[width=9cm,height=8cm,angle=0]{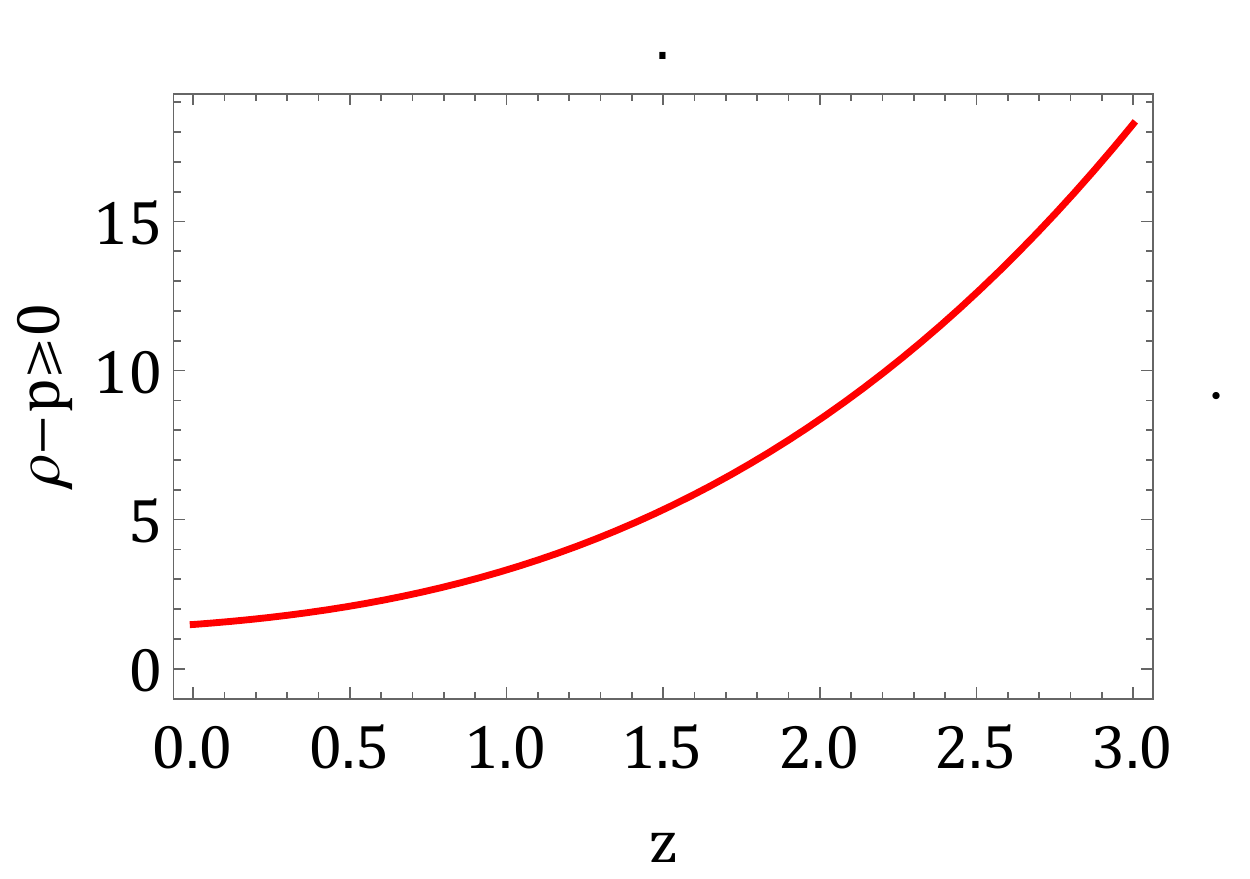}
	\caption{There are the three types of energy conditions (a) Weak energy condition $p+\rho \geq 0$ (b) Strong energy condition $3p+\rho \geq 0 $ and (c) Dominant energy condition $\rho -p \geq 0$. It is interesting that  all the three conditions are being satisfied here despite negative pressure at present.  Here $\mu = 130.$}
\end{figure}

\section{Distant Modulus, Apparent Magnitude and Estimation of  $H_0$ from their Data Set:}
On the basis of the observations of some $42$ low red shift super nova's, Supernova Cosmology Project and Supernova search team {[ \cite{4},\cite{5}]} headed by noble laureates S. Permutter and M. Riess, found that the distance modulus of these standard candles are more than what was predicted theoretically by standard FLRW model. There after,
Some more data set such as Union 2.1 compilation \cite{17}, Pantheon Pan-STARRS1 data set \cite{92} and more latest \cite{93} $-$ \cite{95} predict that due to higher values of distance modulus and apparent magnitudes of supernova's, our universe is accelerating. \\
The distance modulus $ \mu(z) $ is defined by \cite{29}
\begin{equation}\label{27}
	\mu(z)= m_b - M = 5 Log D_l(z)+\mu_{0},
\end{equation}
where $ m_b $ and $ M $ are the apparent and absolute magnitude of the standard candle respectively. The luminosity distance $ D_l(z) $  and nuisance parameter
$ \mu_0 $ are defined by 
\begin{equation}\label{28}
\mu_0= 25+5 Log \big(\frac{c}{H_0} \big),
and    
\end{equation}
\begin{equation}\label{29}
D_l(z)=(1+z)H_0 \int_0^z{\frac{1}{H(z*)} dz*},
\end{equation}
respectively.\\

In this section we will estimate present value of Hubble parameter $H_0$, from the two types of data set (a) $ 580$ distance modulus of  supernova's at the different red shift in the range $ 0\leq z \leq 1.5 $ (b)  $66$ Pantheon apparent magnitude $m_b$ data's consisting of the latest compilation of SN Ia $40$ binned plus 26 high redshift  data's in the  range $0.014 \leq z \leq 2.26 $.\\

\subsection{Estimation by Data Set (a):}
We compare the data set values from those obtained theoretically from Eqs. (\ref{27}) $-$ (\ref{29}) by forming the following chi squire functions of the present value of Hubble parameter  $H_0$. 
\begin{equation}\label{30}
\chi^{2}(H_{0}) = \frac{1}{579}  \sum\limits_{i=1}^{580}\frac{[\mu th (z_{i},H_{0}) - \mu ob(z_{i})]^{2}}{\sigma {(z_{i})}^{2}},
\end{equation}

From Eq. (\ref{30}), we find minimum $\chi^2$ for the set of  $H_0$'s in the range (65$-$75). We obtain $\{H_0\to 69.7171\}\}$ for $\chi^2= \{0.97323\}$. This may be called a very good fit.  The fit will be more clear from the following error bar and likelihood plots in the figure $6$.\\

Figure $6(a)$ describes the growth of  distant modulus $\mu$ over red shift $ `z'$. The distant modulus  is increasing function of red shift which means that in the past distant modulus was more. It is gradually decreasing over time. The figure also shows that theoretical graph passes near by through the dots which are observed values at different red shifts. Vertical lines are error bars. Figure $6(b)$ is likelihood probability curve for Hubble parameter. Estimated value $H_0=69.1434$ is at the peak.
\begin{figure}[H]
(a)	\includegraphics[width=9cm,height=8cm,angle=0]{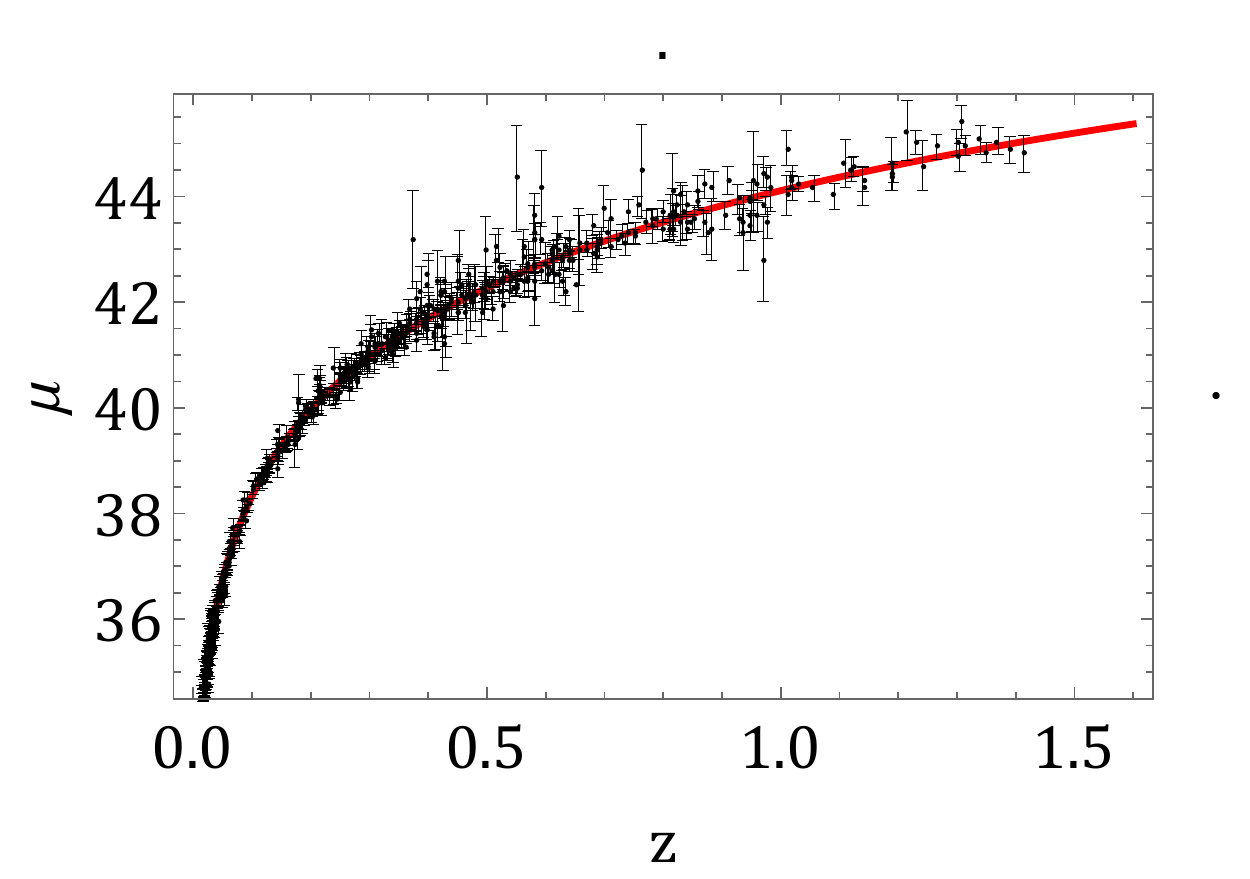}
(b) \includegraphics[width=9cm,height=8cm,angle=0]{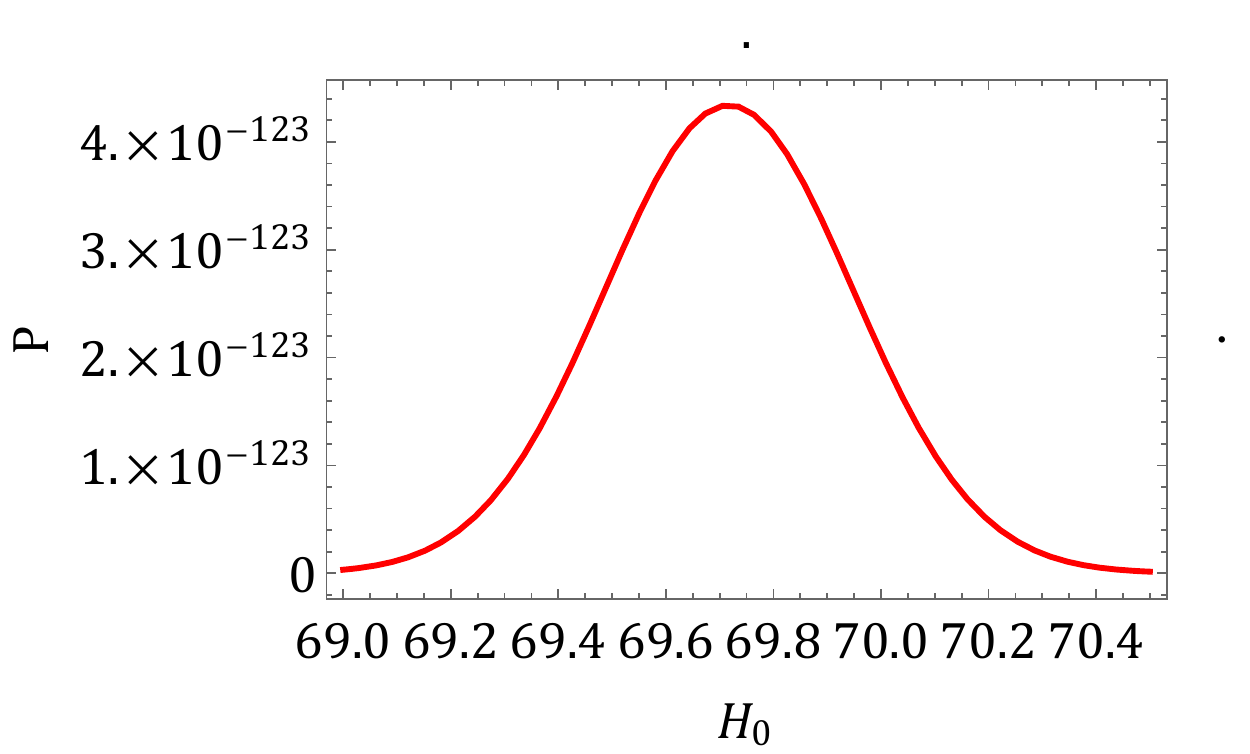}
\caption{Figure (a) is the error bar plots for  distant modulus $\mu$ over red shift $ `z'$. Figure (c) is likelihood probability curve for Hubble parameter. Estimated value $H_0=69.7171$ is at the peak.}
\end{figure}

\subsection{Estimation by Data Set (b):} The data set  (b) is comprised of 66 Pantheon apparent magnitude $m_b$ data's in which $40$ are SN Ia  binned and $26$ high redshift  data's in the range $0.014 \leq z \leq 2.26 $. The literature describes that the absolute magnitudes of a standard candles are more or less same. It  is found \cite{29} that the absolute magnitude of a supernova is obtained as $ M = -19.09$, so we can convert all the apparent magnitude $m_b$ data's into distant modulus data's and vice-versa by adding or subtracting 19.09 into them. We may define expression for $m_b$ as follows:
\begin{equation}\label{31}
m_b =  5 Log[ (1+z) c \int_0^z{\frac{1}{H(z*)} dz*}]+ 5.91	
\end{equation}   
We compare the data set values from those obtained theoretically from Eq. (\ref{31})  by forming the following chi squire functions of the present value of Hubble parameter  $H_0$. 
\begin{equation}\label{32}
\chi^{2}(H_{0}) =\frac{1}{65}  \sum\limits_{i=1}^{66}\frac{[m_{b} th (z_{i},H_{0}) - m_{b} ob(z_{i})]^{2}}{\sigma {(z_{i})}^{2}},
\end{equation}
From Eq. (\ref{32}), we find minimum $\chi^2$ for the set of  $H_0$'s in the range (65$-$75). We obtain $\{H_0\to 78.8906\}\}$ for $\chi^2= \{1.1144\}$. As it is nearer to one so it will also be called a good fit statistically. The fit will be more clear from the following error bar and livelihood plots in the figure $7$.\\

Figure $7(a)$ describes the growth of  apparent magnitude $m_b$ over red shift $ `z'$. The apparent magnitude $m_b$ is increasing function of red shift which means that in the past distant modulus was more. It is gradually decreasing over time. The figure also shows that theoretical graph passes near by through the dots which are observed values at different red shifts. Vertical lines are error bars. Figure $7(b)$ is likelihood probability curve for Hubble parameter. Estimated value $H_0 = 78.8906$ is at the peak.

\begin{figure}[H]
(a)	\includegraphics[width=9cm,height=8cm,angle=0]{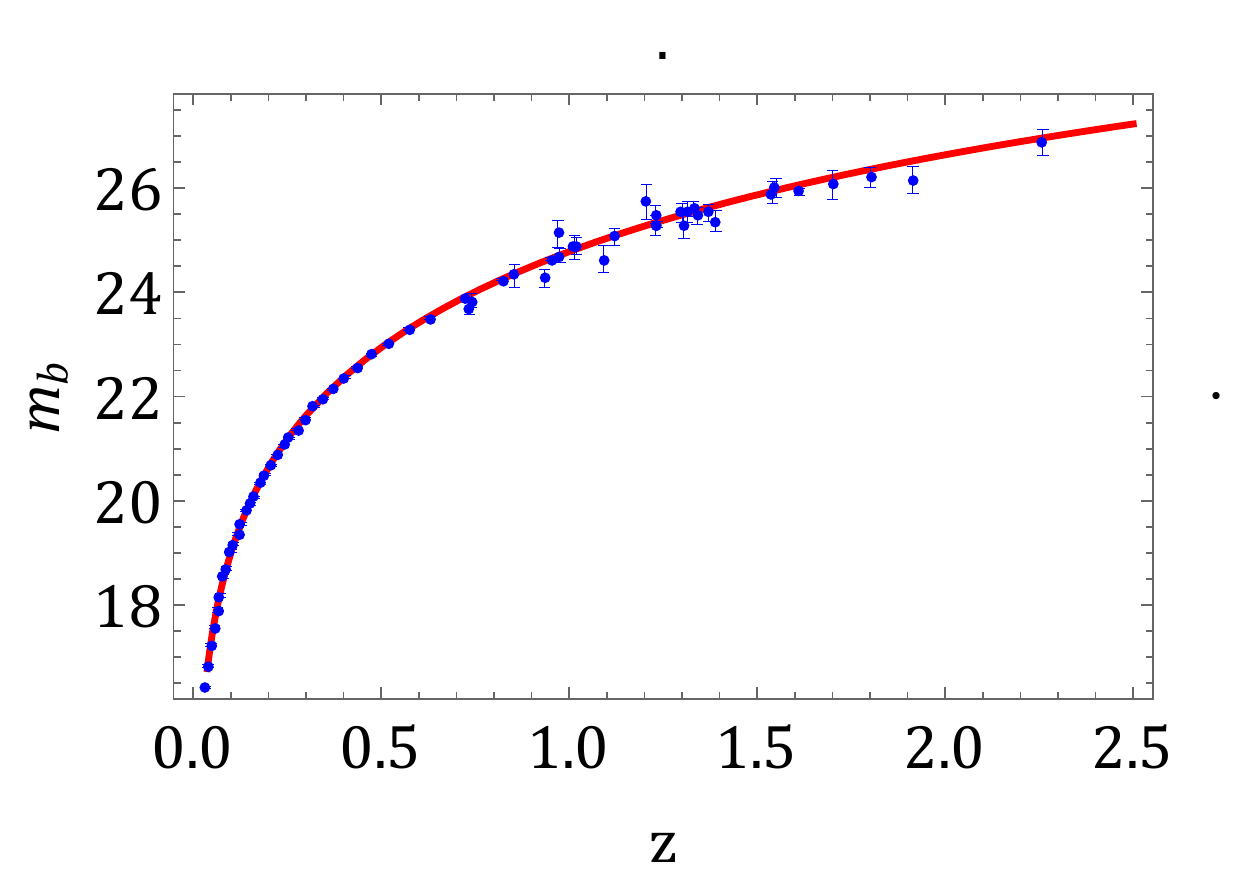}
(b) \includegraphics[width=9cm,height=8cm,angle=0]{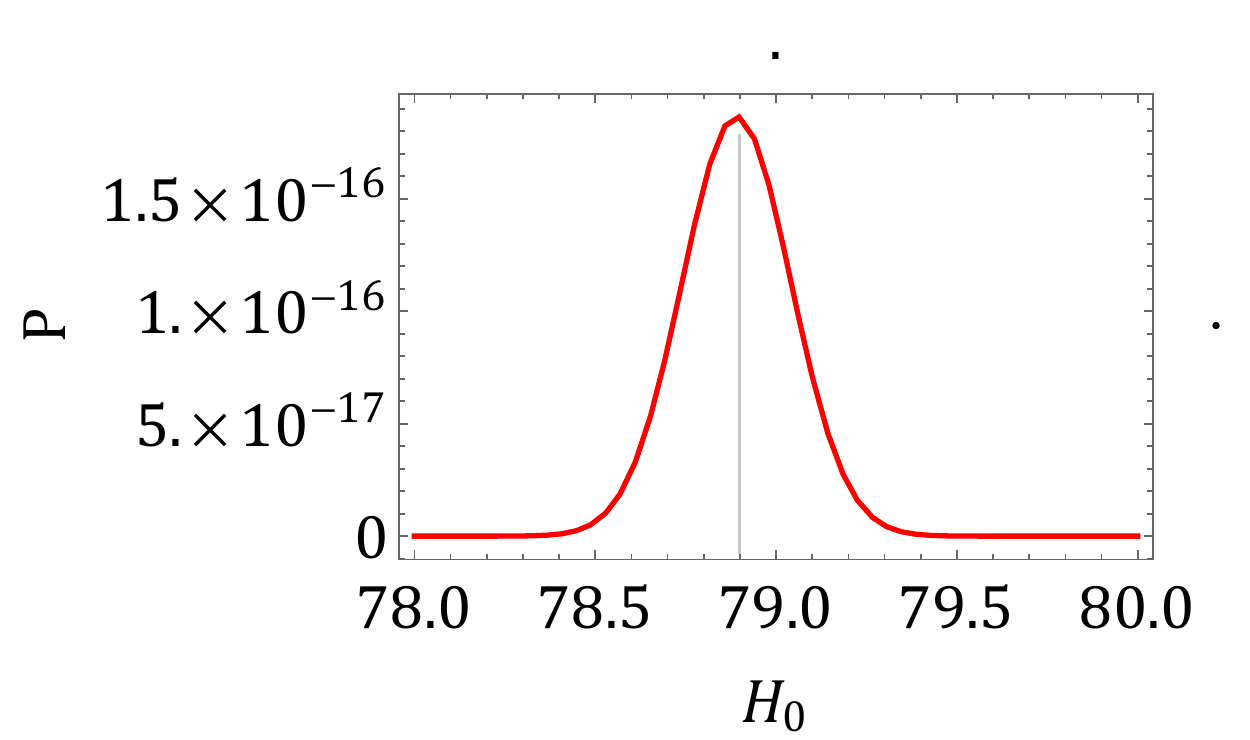}
\caption{Figure (a) is the error bar plots for apparent magnitude $m_b$ over red shift $ `z'$. Figure (b) is likelihood probability curve for Hubble parameter. Estimated value $H_0 = 78.8906$ is at the peak.}
\end{figure}
\section{Estimations  of $H_0$ by combing the data sets:} In this section we will combine the various data sets used earlier. It is found that if we combine Hubble $46$ data set with data set (a), we get a data set of $657$ data's. this gives the estimated value of $H_0$ as $H_0 = 69.367$ for minimum chi squire $\chi2 = 0.864739$. Similarly on combining data set (a) and (b),  we get a data set of 646 data's. This gives the estimated value of $H_0$ as $ H_0 = 78.5398$ for minimum chi squire $\chi2 =1.10473$. last if we combine all the three data sets,  we get a data set of $657$ data's. this gives the estimated value of $H_0$ as $H_0 = 77.914$ for  $\chi2 = 1.04471$. We present the following table to display all of our statistical findings.
\subsection{Time versus Red Shift, Age of the Universe and Transitional times:}
We can calculate the time of any event from the red shift through the following transformation

\begin{equation}\label{33}
	(t_0-t_1)=\intop_{t_1}^{t_0}dt=\intop_{a_1}^{a_0}\frac{da}{aH}=\intop_{0}^{z_1}\frac{dz}{(1+z)H(z)},
\end{equation}
where we have used $\frac{a_0}{a}=1+z $ and $ \dot{z}=-(1+z)H.$ $t_0$ and $t_1$ are present and some past time. We note that at present,t=$t_0$ and z=0. 
With the help of expression for Hubble parameter Eq. (\ref{18}), we can plot the graph of time versus red shift relation. This is given in following the figure $8$.\\
The figure $8$ describe time versus red shift transformation. There is a asymptote in the curve which gives the present age of the universe as $H_0 t_0$= 0.9410.
On the basis of the various estimates of $H_0$, the age of the universe comes to $13.5245*10^9$ yrs and  $12.0854 \times 10^9$ yrs in our model. The transition time where the universe inters the accelerating phase is also calculated as $4.10953 \times 10^9$ yrs and  $3.67228 \times 10^9$ yrs as from now.

\begin{table}[H]
\caption{ Statistical estimations for Hubble parameter $H_0$}
\begin{center}
	\begin{tabular}{|c|c|c|c|}
		\hline
		\\
		Datasets &~	 $H_{0}$~ &~   $\chi^2$  ~ \\
		\\	
		\hline
		\\
		$OHD$  &~ $ 68.95 $ ~ &~ $0.5378~ $ \\
		\\	
		\hline
		\\
		D.M	$\mu$ &~  $69.7171$	 ~&~ $0.97323~ $\\
		
		\\	
		\hline
		\\
		A.M	$m_b$ &~  $78.8906$	 ~ &~ $1.1144~ $\\
		\\	
		\hline
		\\
		$OHD$ +	D.M	$\mu$ &~  $69.367$	 ~ &~  $ 0.864739~$\\
		\\	
		\hline
		\\	
		D.M	$\mu$ + A.M	$m_b$ &~  $78.5398$	  ~&~ $ 1.10473~$\\
		\\	
		\hline
		\\
		$OHD$ + D.M	$\mu$ + A.M	$m_b$	 &~  $77.914$	 ~&~  $ 1.04471~$\\
		\\
		\hline			
	\end{tabular}
\end{center}
\end{table}

\begin{figure}[H]
\includegraphics[width=9cm,height=8cm,angle=0]{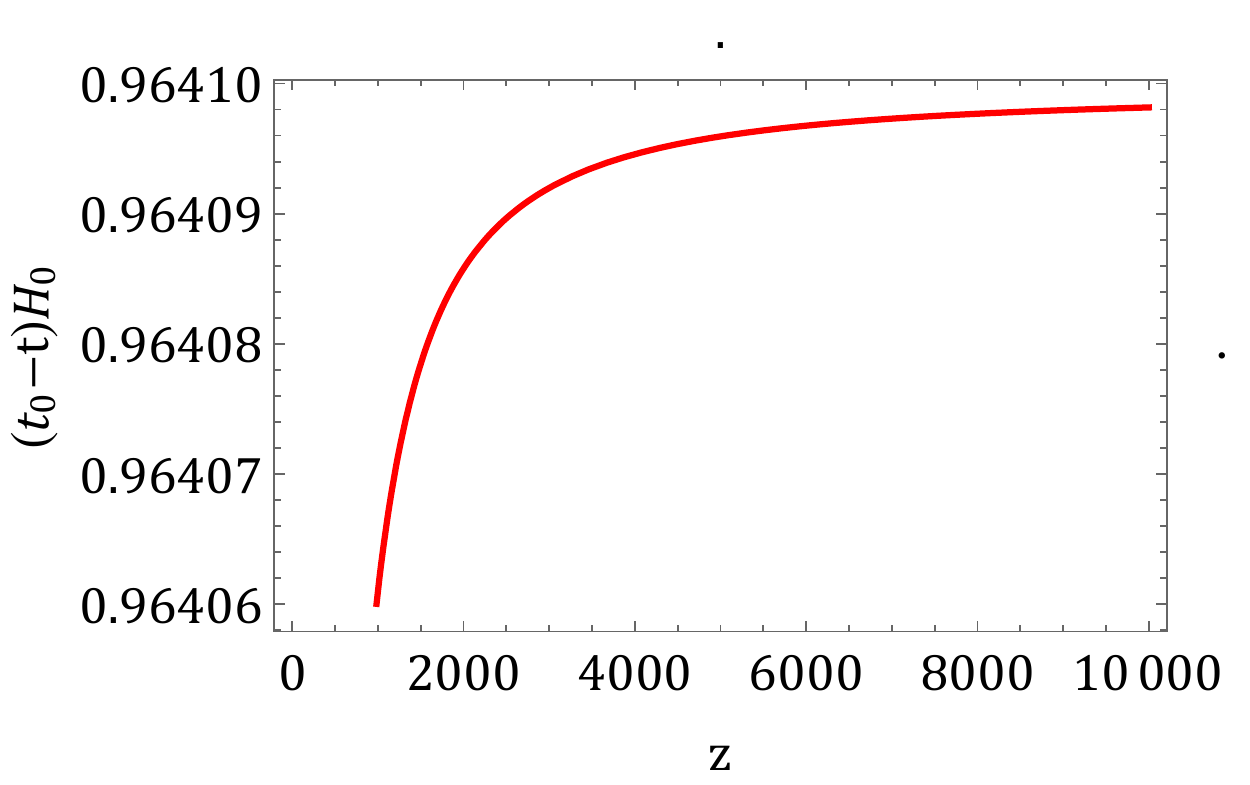}
\caption{ The figure describe time versus red shift transformation. There is a asymptote in the curve which gives the present age of the universe as $H_0 t_0$= 0.9410. }
\end{figure}
\section{Conclusion:} In  a brief nutshell, we say that we have developed a FLRW type cosmological model of the universe in $f(R,T)$ gravity theory which deviates from its original features as per the needs of the present day observations. The main findings of our model are stated point wise.
\begin{itemize}
\item  The model displays transition from deceleration in the past to the acceleration at the present. The three parameters namely the deceleration parameter $q$, pressure $p$ and equation of state parameter $\omega$ carries this transition.\\

\item The interesting part in the paper is that the  transition red shift is found similar in all the three parameters stated above in the item (i).
\item  Energy conditions in the model are evaluated and discussed. Interestingly, the model satisfies all the three  weak, strong and Dominant energy conditions despite the presence of the negative pressure at present.
\item We have considered the simplest form of $f(R,T)$ as $f(R,T)= R+ \Lambda T$ and estimated $ \mu = 8 \pi \lambda =130.$
\item We will estimate present value of Hubble parameter $H_0$, from the three types of data set (a)  580 distance modulus of  supernova's at the different red shift in the range $ 0\leq z \leq 1.5 $ (b)  66 Pantheon apparent magnitude $m_b$ data's consisting of the latest compilation of SN Ia 40 binned plus $26$ high redshift  data's in the  range $0.014 \leq z \leq 2.26 $ (c) latest compilation of  $46$( OHD)  observational Hubble  data set.The out come of these is displayed in Table $1$.
\item We have formulated apparent magnitude of a standard candle (see Eq. (31)).
\item With the help of the various error bar plots and likelihood plots values of chi-squire, we tried to show how much our theoretical result tallies with  the observational ones.
\item We have calculated the age of the universe as per our model. one of the estimation is $13.5$ billion yrs. This value is at par with the observed one. Similarly we have also estimated Hubble constant at present  as $68.95$ This value is  also at par with the present observed values\cite{20} $-$ \cite{23}.

\item  If we look at the table$-$ 1, we find that the present Hubble parameter has variations in its value   on the basis of estimations by the three data sets and their combinations. It is observed that the $H_0\simeq 69$-$70$ for OHD and SNIa 580 distance modulus data, whereas when we use the pantheon data set, $H_0\simeq 78$-$79$. There is considerable difference in these values. We recall that we also face the same issue  when we empirically find the present value of the Hubble parameter using calibrated distance ladder techniques and measurements of the cosmic microwave background. The former value comes to $H_0\simeq 73$ whereas the CMB based value is $H_0\simeq 68$. This is often given the name Hubble Tension.
\end{itemize}
The authors are confident that readers and researchers will find our work valuable in the final analysis.

\section*{Acknowledgments}
The authors (A. Pradhan \& G. K. Goswami ) are grateful for the assistance and facilities provided by the University of Zululand, South Africa during a visit where a part of this article was completed. Sincere thanks are due to the anonymous reviewer for his constructive comments to enhance the quality of the paper.

\end{document}